\documentclass[usegraphicx,usenatbib]{mn2e}
\newcommand{\aap}{A\&A}
\newcommand{\apj}{ApJ}
\newcommand{\apjl}{ApJ}
\newcommand{\apjs}{ApJS}
\newcommand{\araa}{ARA\&A}
\newcommand{\mnras}{MNRAS}
\newcommand{\nat}{Nat}

\newcommand\pasp{\ref@jnl{PASP}}

\usepackage{amsmath}
\usepackage{amssymb}
\usepackage{graphicx}
\usepackage{color}
\usepackage{nicefrac}
\usepackage{footnote}

\graphicspath{{./figures/}}

\newcommand{\note}[1]{\emph{\textcolor{red}{}}}

\newcommand{\Ms}{{\ensuremath{\mathrm{M}_{\odot} }}}

\newcommand{\Ni}{{\ensuremath{^{56}\mathrm{Ni}}}}
\newcommand{\Fe}{{\ensuremath{^{56}\mathrm{Fe}}}}

\newcommand{\He}{{\ensuremath{^{4} \mathrm{He}}}}

\newcommand{\Hy}{{\ensuremath{^{1} \mathrm{H}}} }
\newcommand{\Ox}{{\ensuremath{^{16}\mathrm{O}}}}
\newcommand{\Ti}{{\ensuremath{^{44}\mathrm{Ti}}}}
\newcommand{\Si}{{\ensuremath{^{28}\mathrm{Si}}}}
\newcommand{\Mg}{{\ensuremath{^{24}\mathrm{Mg}}}}
\newcommand{\Cx}{{\ensuremath{^{12}\mathrm{C}}}}

\newcommand{\Cr}{{\ensuremath{^{48}\mathrm{Cr}}}}
\newcommand{\Ca}{{\ensuremath{^{40}\mathrm{Ca}}}}
\newcommand{\Ar}{{\ensuremath{^{36}\mathrm{Ar}}}}
\newcommand{\Sx}{{\ensuremath{^{32}\mathrm{S}}}}
\newcommand{\Nx}{{\ensuremath{^{14}\mathrm{N}}}}
\newcommand{\Ne}{{\ensuremath{^{20}\mathrm{Ne}}}}

\newcommand{\CASTRO}{\texttt{CASTRO}}
\newcommand{\KEPLER}{\texttt{KEPLER}}

\title[]{Low-Energy Population~III Supernovae and the Origin of Extremely Metal-Poor Stars}

\author[Chen et al.]{Ke-Jung Chen$^{1,2,3}$\thanks{EACOA Fellow, E-mail: ken.chen@nao.ac.jp},
	     Alexander Heger$^{4,5,6}$,  
	     Daniel J. Whalen$^{7}$, 
	     Takashi J. Moriya$^{1}$,  
	     \newauthor
	     Volker Bromm$^{8}$, 
	     and S.~E. Woosley$^{3}$	 \\
$^{1}$  Division of Theoretical Astronomy, National Astronomical Observatory of Japan, Tokyo 181-8588, Japan     \\
$^{2}$ Institute of Astronomy and Astrophysics, Academia Sinica,  Taipei 10617, Taiwan \\
$^{3}$ Department of Astronomy \& Astrophysics, University of California, Santa Cruz, CA 95064, USA \\
$^{4}$  School of Physics and Astronomy, University of Minnesota, Minneapolis, MN 
55455, USA \\
$^{5}$  Monash Centre for Astrophysics, School of Mathematical Sciences, Monash University, 
Victoria 3800, Australia \\
$^{6}$   Joint Institute for Nuclear Astrophysics, University of Notre Dame Notre Dame, IN 
46556 USA \\
$^{7}$  Institute of Cosmology and Gravitation, Portsmouth University, Portsmouth, UK \\
$^{8}$  Department of Astronomy, University of Texas, Austin, TX 78712, USA}

\begin{document}

\maketitle
\topmargin-1cm

\begin{abstract}

Some ancient, dim, metal-poor stars may have formed in the ashes of the first supernovae.  If their chemical abundances can be reconciled with the elemental yields of specific Pop III explosions, they could reveal the properties of primordial stars.  But multidimensional simulations of such explosions are required to predict their yields because dynamical instabilities can dredge material up from deep in the ejecta that would otherwise be predicted to fall back onto the central remnant and be lost in one-dimensional (1D) models.  We have performed two-dimensional (2D) numerical simulations of two low-energy Pop III supernovae, a 12.4 \Ms\ explosion and a 60 \Ms\ explosion, and find that they produce elemental yields that are a good fit to those measured in the most iron-poor star discovered to date, SMSS J031300.36-670839.3 (J031300).  Fallback onto the compact remnant in these weak explosions accounts for the lack of measurable iron in J031300 and its low iron-group abundances in general. Our 2D explosions produce higher abundances of heavy elements (atomic number $Z >$ 20) than their 1D counterparts due to dredge-up by fluid instabilities. Since almost no \Ni\ is ejected by these weak SNe, their low luminosities will prevent their detection in the near infrared with the {\it James Webb Space Telescope} and future 30-meter telescopes on the ground. The only evidence that they ever occurred will be in the fossil abundance record.

\end{abstract}


\section{Introduction}

The formation of the first stars marked the end of the cosmic Dark Ages and began the transformation of the universe from a cold, dark featureless void into the hot, transparent cosmic web of galaxies we observe today \citep{ABN2002,BCL2002,wan04,kitayama2004,
alvarez2006,abel2007}. Population III (or Pop III) stars were also the first great nucleosynthetic engines of the universe, expelling large quantities of the heavy elements needed for the later formation of planets and life \citep[e.g.,][]{byh03,ky05,get07,karlsson2013}.  They may also be the origins of the supermassive black holes found in most massive galaxies today \citep[e.g.,][]{vhm03,milos2009a,alvarez2009,pm11,jlj12a,agarw12,jet13,latif13c,jet14,smidt16,tyr16,hle16}.   

Modern cosmological simulations suggest that Pop III stars formed at $z \sim$ 20 in small pre-galactic structures known as minihalos, with masses $\gtrsim 10^5$ \Ms\ \citep{fsg09,fg11,
dw12,glov13}.  The original models found that Pop III stars were 100 - 500 \Ms\ and formed in isolation, one star per halo.  Newer simulations indicate that Pop III stars may also have formed in binaries \citep{turk2009,stacyB2013} or in small clusters \citep{stacy2010,clark11,greif2011,
greif2012,sbl16}.  The most recent simulations of ionizing UV breakout from Pop III protostellar disks show that accretion onto the star may be terminated at only a few tens of solar masses \citep{hoso2011,stacy2012,hirano2014,stacy2014}, although 60 - 1000 \Ms\ stars are still possible \citep{hir15}.  However, in spite of their extreme luminosities \citep{BKL2001,
schaerer2002}, individual Pop III stars will not be visible even to the coming generation of 30-m telescopes \citep[but see][]{rz12}.

Direct detections of Pop III SNe could constrain the properties of their progenitors. Non-rotating Pop III stars from 10 - 40 \Ms\ die as core-collapse supernovae (CC SNe) and 90 - 260 \Ms\ stars can explode as pair-instability (PI) SNe \citep{barkat1967,glatzel1985,heger2002,heger2010,chatz2012,cwc13,cw14a,chen2014b}. A recent arc of radiation hydrodynamical simulations has shown that PI SNe will be visible in the near infrared (NIR) at $z \sim$ 15 - 20 to the {\it James Webb Space Telescope} ({\it JWST}), the Wide-Field Infrared Survey Telescope (WFIRST) and the Thirty-Meter Telescope \citep{kasen11,cooke12,hum12,pan12a,wet12a,wet13d,wet12b,wet12d,ds13,ds14,smidt14a} and that CC SNe and gamma-ray bursts (GRBs) will be visible out to $z \sim$ 10 - 15 \citep[e.g.,][]{wet08c,moriya10,tet12,wet12c,wet12e,tet13,mes13a,smidt13a,ds14,magg16}.

Another path to constraining the Pop III initial mass function (IMF) is to look for the chemical fingerprint of the first SNe on later generations of stars. The basic idea is that if stars of lower and lower metallicity can be found, a few will be encountered that must have formed in the debris of the very first SNe. If these chemical patterns can then be matched to the elemental yields of specific Pop III explosions (or a well-defined population of them) they could reveal important clues about the masses of the first stars. These abundances, known collectively as the fossil abundance record, have been the focus of several past and ongoing surveys of ancient, dim metal-poor stars in the Local Group \citep[e.g.,][]{Cayrel2004,beers2005,frebel2005,Lai2008} and of low-mass galaxies \citep{cooke14,
venn2014} \citep[see also][]{johnson2014}.  The elemental patterns found in metal-poor stars to date suggest that many were enriched by 10 - 40 \Ms\ Pop III stars \citep{umeda2003,heger2010,candace2010,plac16}, although there is now evidence of the odd-even nucleosynthetic imprint of PI SNe in the metal-poor star J001820 \citep{aoki14} \citep[see also][]{tumlinson2006,karlsson2008}.  It has now also been found that the stars in some ultra-faint dwarf galaxies (UFDs) may have been chemically enriched by just a single low-mass Pop III SN \citep{sim10,fb12}.

The most iron-poor star found to date, SMSS J031300.36-670839.3 (J031300), by \citet{keller2014, Bes2015}, has an [Fe/H] ratio below $10^{-7}$.  This unusually low iron abundance has been attributed to fallback onto a central black hole (BH) in the weak SN of a 60 \Ms\ star.  In these weak and therefore faint events, the innermost shell of ejecta with elements such as Si and Fe likely falls back onto the central compact object during the explosion.  This process could explain the near absence of iron group elements in J031300.  Similar yields have been reported for low-energy ($\lesssim$ 0.5 B, where 1 B $=$ 10$^{51}$ erg) explosions of less massive progenitors \citep[see][]{marass2014,Ishigaki2014,
Tak2014}.  Such events are considered to be low-energy because they have energies that are either well below the average for CC SNe \citep[$\sim$ 0.9 B;][]{kasen2009} or, as in the case of the 60 \Ms\ progenitor discussed above, have low energies per unit mass.

But mixing between adjacent shells of elements during the explosion could offset fallback in the final elemental yields of a Pop III SN because hydrodynamical instabilities can dredge elements deep in the ejecta up to a higher mass coordinate, beyond the grasp of the BH or neutron star (NS).  One-dimensional (1D) explosion models, even those with extensive nuclear reaction networks, therefore cannot predict the chemical yields of low-energy Pop III SNe from first principles.  Mixing on small scales in the explosion must also be understood to predict the metallicities of the stars that form in the debris of the SN because it sets the initial conditions for mixing, cooling and collapse of gas on larger scales in the intergalactic medium (IGM).  Mixing in Pop III CC and PI SNe has previously been studied by \citet{joggerst2009,joggerst2011,
chen2011,chen2014b} and on larger, cosmological scales by \citet{wise2008,greif2010,
jeon2014,chiaki2015,che2015,ritter2015, sluder2015}. 

We have now performed the first two-dimensional (2D) simulations of low-energy Pop III SNe with the \CASTRO{} code to determine if their elemental yields can be a good match to those of J031300.  We describe our progenitor stars and \CASTRO{} models in Section 2.  Mixing, elemental yields and observational signatures are examined in Sections 3 and 4, and we conclude in Section 5.

\section{Numerical Method}

Our two fiducial models are the core-collapse explosions of 12.4 \Ms\ and 60 \Ms\ Pop III stars, designated Z12 and Z60, respectively. They were chosen from a large database of 1D Pop III SN models \citep{heger2010} because their elemental yields were in good agreement with those of J031300.  These stars were first evolved from the zero-age main sequence to the onset of core collapse with the 1D stellar evolution code \KEPLER{} \citep{kepler,heger2001} as described in \citet{heger2010}.  The explosions were then triggered with a linear momentum piston with energies of 0.6 B and 1.8 B, respectively.  

After the launch of the piston, a shock forms at the base of the silicon burning shell. The postshock temperature drops below $10^9$ K within ten seconds and most nuclear burning ends.  At this point profiles for the ejecta, the star and the surrounding medium are mapped onto a 2D cylindrical grid in \CASTRO{}. The mapping preserves conservation of quantities such as mass and energy over the large range of spatial scales in the star that must be resolved to properly evolve the flow, such as the burning layer on scales of 10$^8$ cm to breakout from the star at 10$^{12}$ - 10$^{13}$ cm \citep{chen2013}. 

\begin{figure}
\begin{center}
\includegraphics[width=\columnwidth]{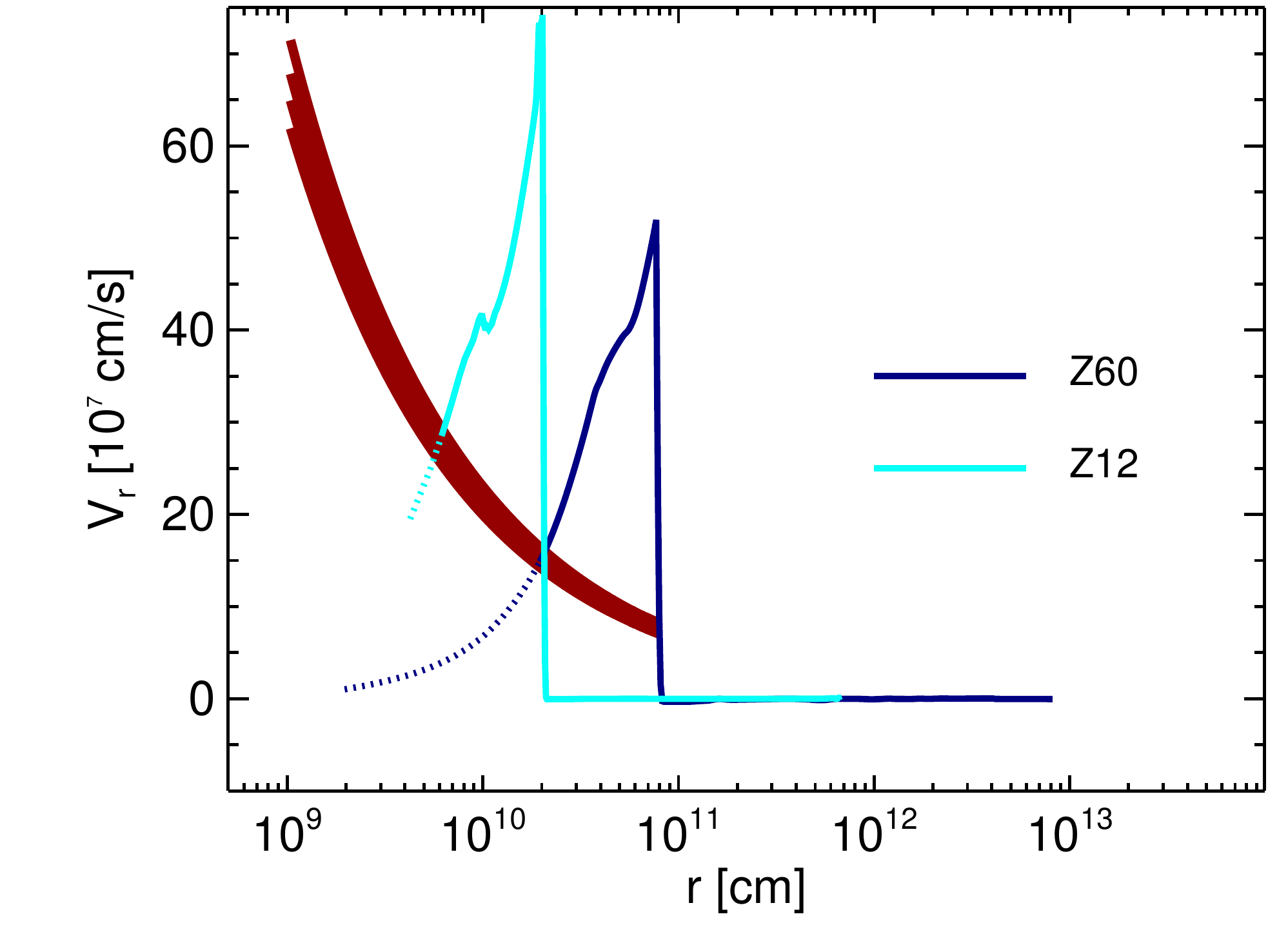} 
\caption{1D velocity profiles at the beginning of the \CASTRO{} runs.  The red stripe shows escape 
velocities for 1.5 - 2 \Ms\ neutron stars.  Ejecta (dotted lines) with velocities below this band may not escape.  
\label{fig:initv}}
\end{center}
\end{figure}

\subsection{\CASTRO}

\CASTRO{} is a multidimensional adaptive mesh refinement (AMR) code \citep{ann2010,zhang2011} with an unsplit piecewise-parabolic method (PPM) hydrodynamics scheme \citep{woodward1984}. It has a realistic equation of state for stellar interiors from \citet{timmes2000} that includes contributions from both degenerate and non-degenerate relativistic and non-relativistic electrons, electron-positron pairs, ions (which are treated as an ideal gas), and photons.  To track the mixing of elements, we follow the advection of 15 isotopes, \Hy, \He, \Cx, \Nx, \Ox, \Ne, \Mg, \Si, \Sx, \Ar, \Ca, \Ti, \Cr, \Fe, \Ni.  The radii of the Z12 and Z60 stars are $7\times10^{11}$ cm and $2\times10^{13}$ cm, respectively. We only simulate a quarter of each star in angle, with outer boundaries in $r$ and $z$ at 10 times the radius of the star.  To resolve fluid instabilities in 2D, we use $256 \times 256$ zones on the root grid and up to eight levels of AMR for an additional factor of up to $2^8 = 256$ in resolution.  This approach yields an effective simulation domain of $65,536 \times 65,536$ zones. 

We adopt refinement criteria based on gradients in density, velocity, and pressure and ensure that the inner core, where most fallback occurs, has the highest resolution.  Reflecting and outflow boundary conditions are set on the inner and outer boundaries in $r$ and $z$, respectively.  The monopole approximation for self-gravity, in which a 1D gravitational potential is constructed from the radial average of the density, is used to calculate gravitational forces everywhere in the AMR hierarchy.  This approximation is valid because the star is nearly spherically symmetric.  The gravity of the NS or BH is included in our models.  In both cases the explosion is evolved until the ejecta reaches the outer boundaries of the grid.

\begin{figure}
\begin{center}
\includegraphics[width=.8\columnwidth]{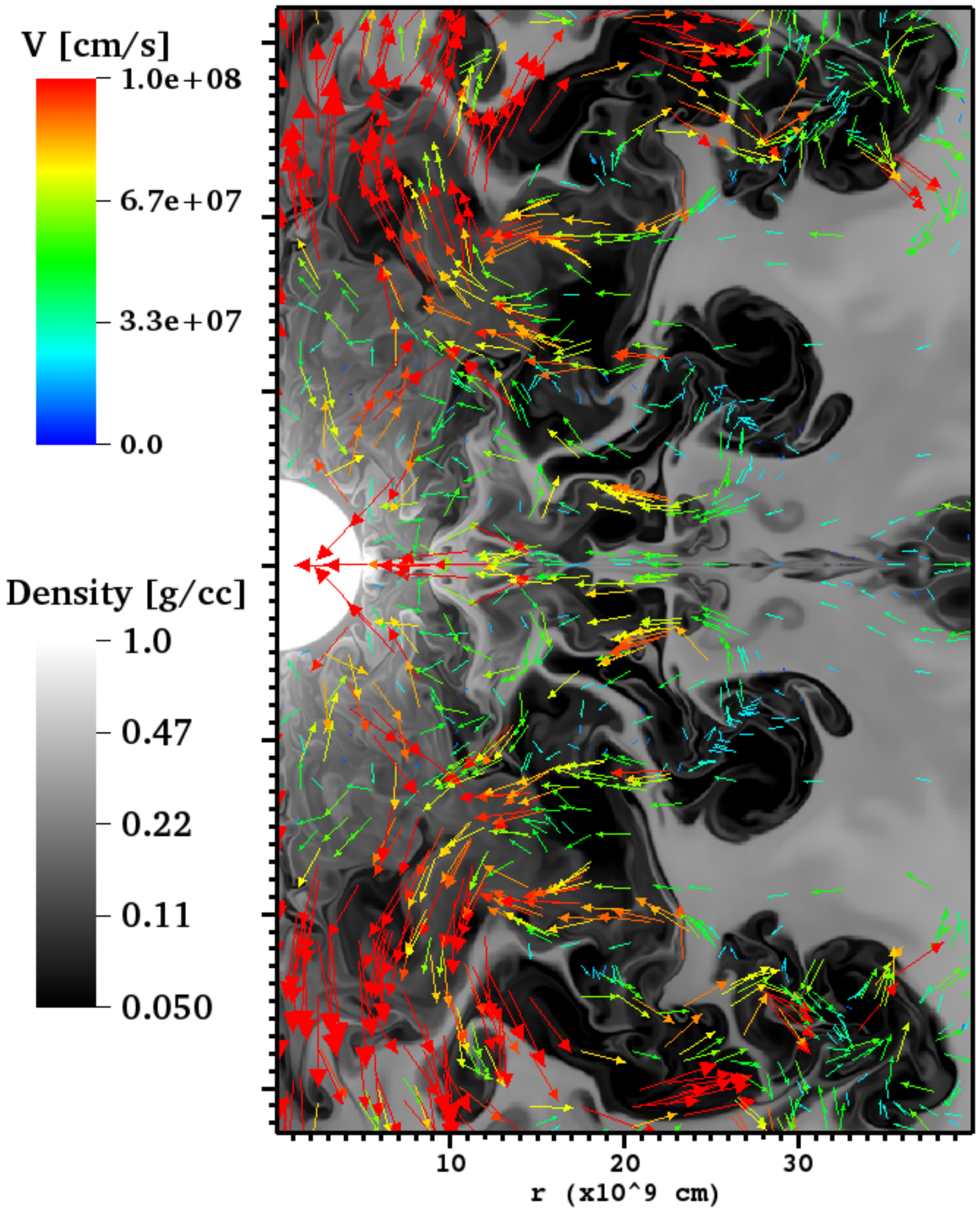} 
\caption{Fallback onto the NS in the Z12 SN at 1,800 sec.  Although the shock has broken out of the star, fallback continues, creating turbulent accretion flows.  Some of the flow eventually settles onto the NS at the center of the white half-oval.  Flows parallel to the axes are visible at the inner boundaries in $z$ and $r$ and are due to pole effects and reflecting boundary conditions there.\label{fig:z12acc}}
\end{center}
\end{figure}

\begin{figure*}
\begin{center}
\includegraphics[width=0.48\textwidth]{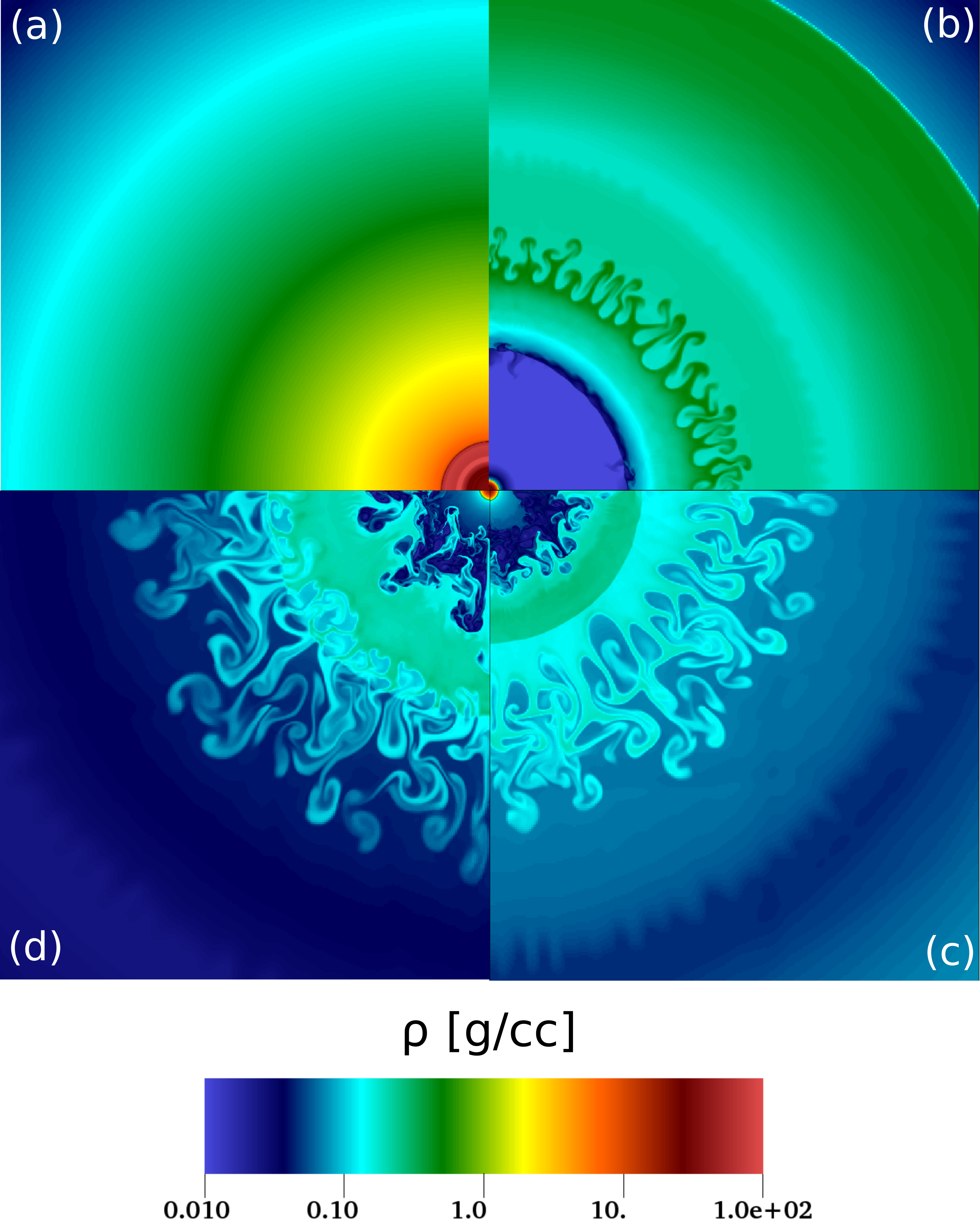} 
\includegraphics[width=0.48\textwidth]{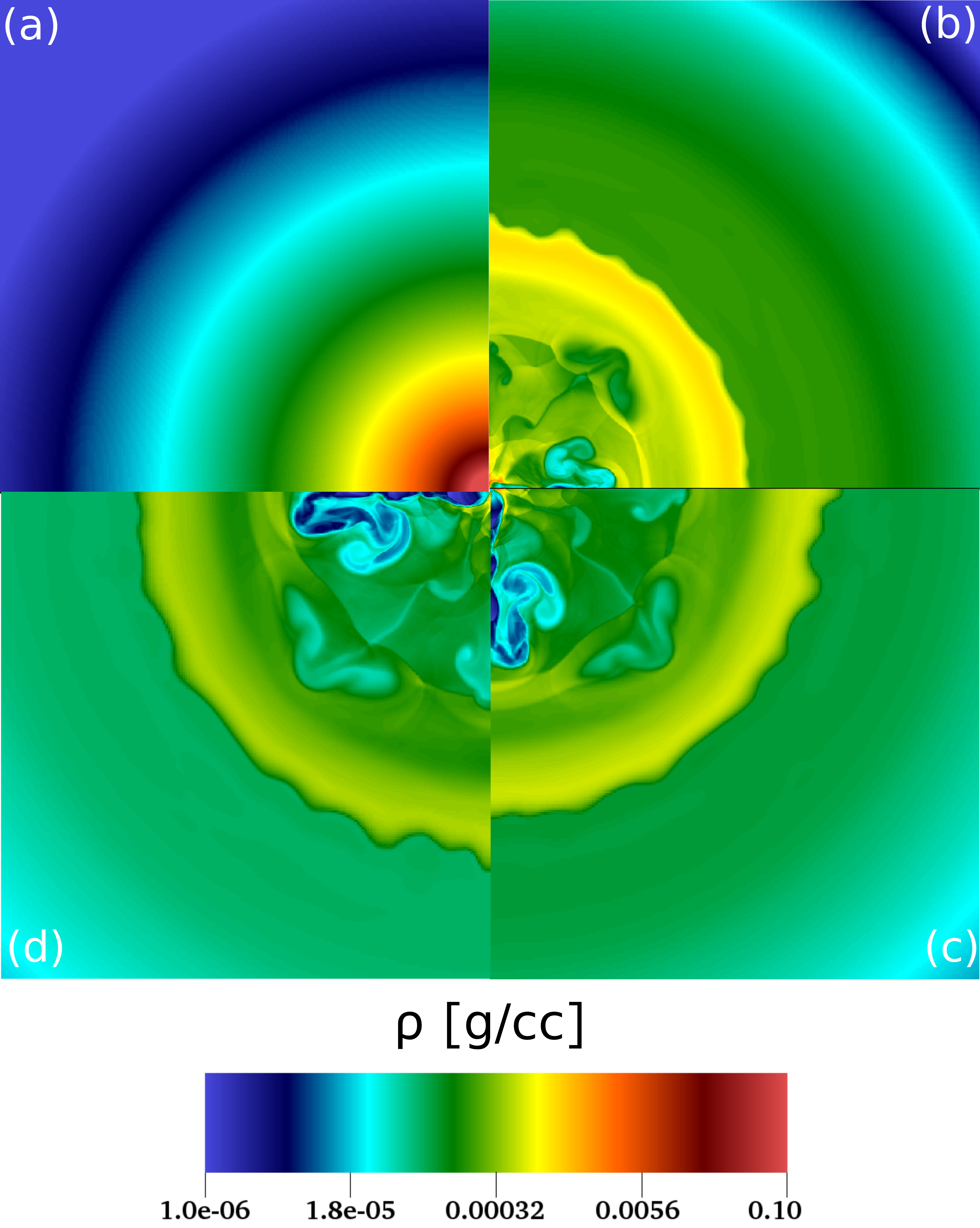} 
\caption{Evolution of fluid instabilities in the Z12 and Z60 models before and after shock breakout.  {\it Left:}  (a) - (d) show the instabilities in Z12 at 0, 340, 1200, and 2300 sec, which are driven by fallback and the formation of the reverse shock.  They first appear at the inner and outer boundaries of the carbon/oxygen burning shells.  The panels are $2 \times 10^{11}$ cm on a side.  {\it Right:}  (a) - (d) show the instabilities in Z60 at 0, 2100, 9200, and 31000 sec.  Here, the instabilities are mainly driven by fallback because mixing due to the reverse shock is much weaker.  These panels are $2 \times 10^{12}$ cm on a side.   
\label{fig:evol}}
\end{center}
\end{figure*}

\section{Explosion Models}

The initial velocity of the shock is lower than those in most CC SNe because of the smaller explosion energies of these two models.  The innermost regions of the ejecta, which are also at the tail of the velocity profile, do not have enough kinetic energy to escape the gravity of the NS or BH, as we show in Figure~\ref{fig:initv}.  The peak velocities are 7.4 $\times$ 10$^8$ cm s$^{-1}$ for Z12 and 5.2 $\times$ 10$^8$ cm s$^{-1}$ for Z60.  We define the escape velocity from the neutron star to be $v_{\mathrm{esc}} = \sqrt{2GM_n/r^2}$, where $G$ the gravitational constant, $M_n$ is the NS mass, and $r$ is the distance to the center of the NS.  Material near the center of the ejecta at velocities below $v_\mathrm{{esc}}$ will fall back onto the NS in 1D models. 

\subsection{Fallback}

In reality, fluid instabilities driven by fallback can dredge up material with velocities $\lesssim v_\mathrm{{esc}}$ and eject it from the star.  We show densities and velocities at the core of the Z12 SN during fallback in Figure \ref{fig:z12acc}.  Rayleigh-Taylor (RT) instabilities form and create accretion streams onto the NS.  The streams become turbulent, as can be seen in the eddies on multiple scales.  Eventually, 0.322 \Ms\ and 4.967 \Ms\ fall back in Z12 and Z60, creating a 1.78 \Ms\ NS and a 6.87 \Ms\ BH, respectively. Although nearly all the iron group elements near the centers of both stars fall back onto the central remnant, more of them escape in our simulations than in 1D models of the same SNe, as we show later. 

\subsection{Shock Breakout / Initial Expansion}

The shock breaks out of Z12 much sooner than from Z60 (at 1,659 sec vs 25,600 sec) because it dies as a compact blue giant rather than a red supergiant, and has a much smaller radius.  The shock crashes out into a diffuse circumstellar medium (CSM) with a density profile $\rho \propto r^{-3.1}$.  This profile was chosen to avoid the formation of a reverse shock after breakout, which would cause additional mixing that is not inherent to the explosions themselves.  The growth of fluid instabilities in Z12 and Z60 before and just after shock breakout is shown in Figure \ref{fig:evol}.  Prominent RT fingers form and mix the \Cx\ and \Ox\ shells and a small fraction of the \Si\ shell in Z12. There is less mixing in the outer shell in Z60 but fallback breaks the spherical symmetry of the inner shells.  

\begin{figure}
\begin{center}
\includegraphics[width=0.48\columnwidth]{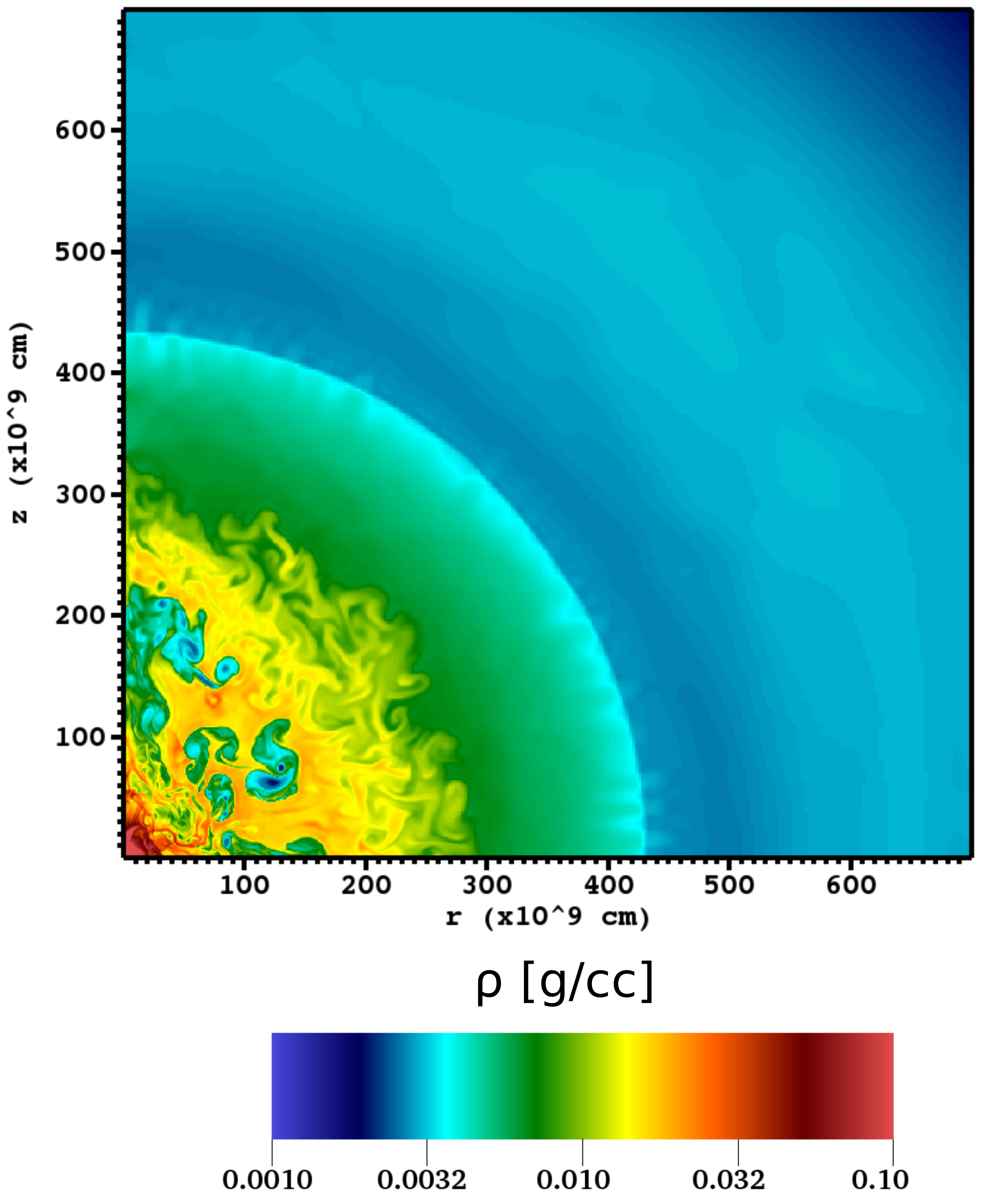} 
\includegraphics[width=0.48\columnwidth]{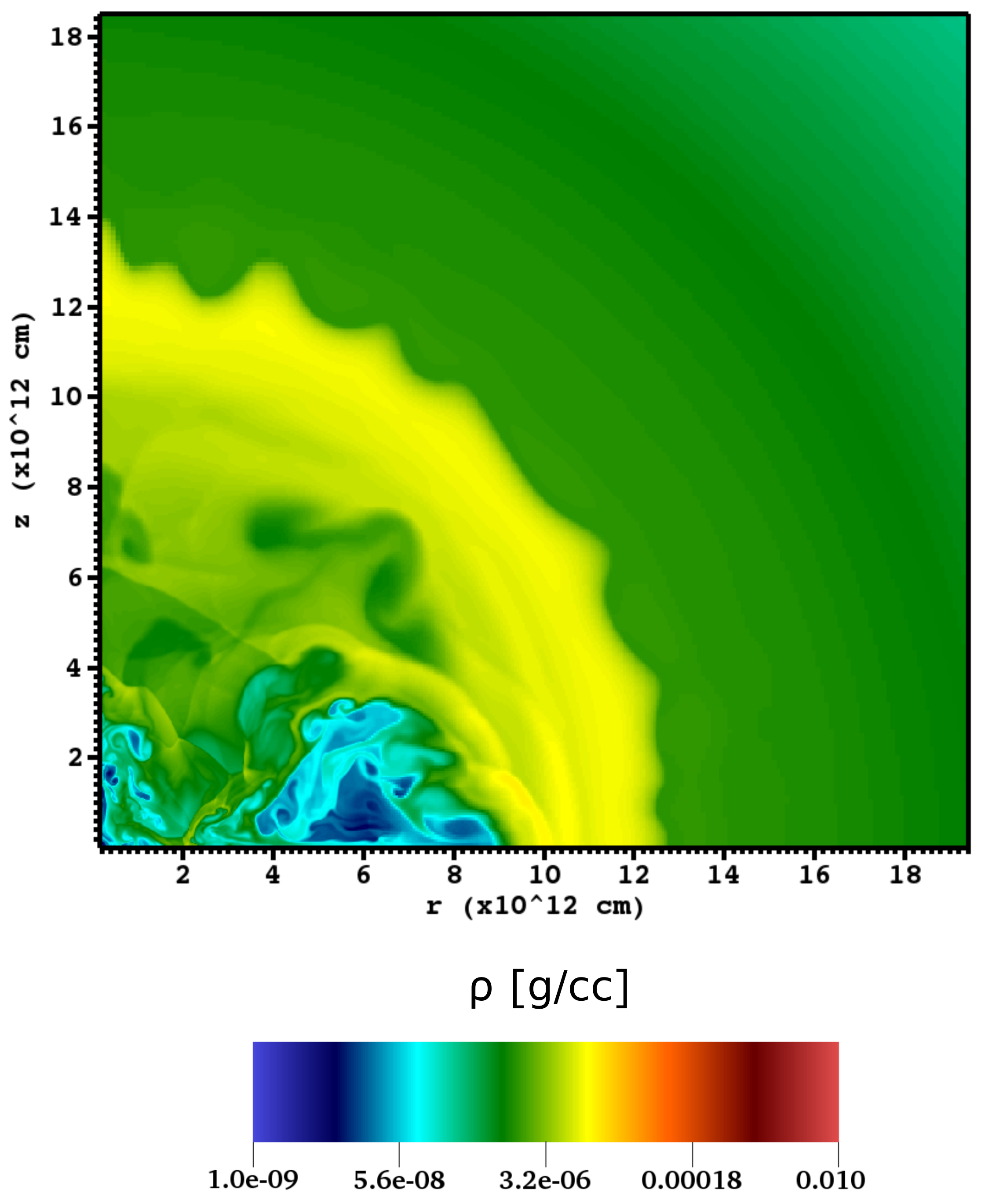} 
\caption{$\it Left$: Fluid instabilities in Z12 at $t\sim$ 44,000 sec.  The reverse shock has reached the center of the star and mixing is now complete, although fallback continues. $\it Right$: Fluid instabilities in Z60 at $t\sim$ 105,000 sec.  The reverse shock has reached the center of the star.  In contrast to Z12, the outer shell is only slightly mixed. \label{fig:break}}
\end{center}
\end{figure}

Ejecta for both models at intermediate times are shown in Figure \ref{fig:break}.  The reverse shock has reached the center of the explosion in both cases.  Mixing driven by reverse shock is still evolving at 44,000 sec in Z12 but is mostly complete in Z60 by 105,000 sec.  Fallback is still happening in both explosions at these times.  The outer shell of the star is heavily mixed in Z12 but not in Z60.  Mixing driven by fallback continues well after shock breakout in both explosions.

\subsection{Homologous Expansion}

\begin{figure}
\begin{center}
\begin{tabular}{cc}
\includegraphics[width=0.5\columnwidth]{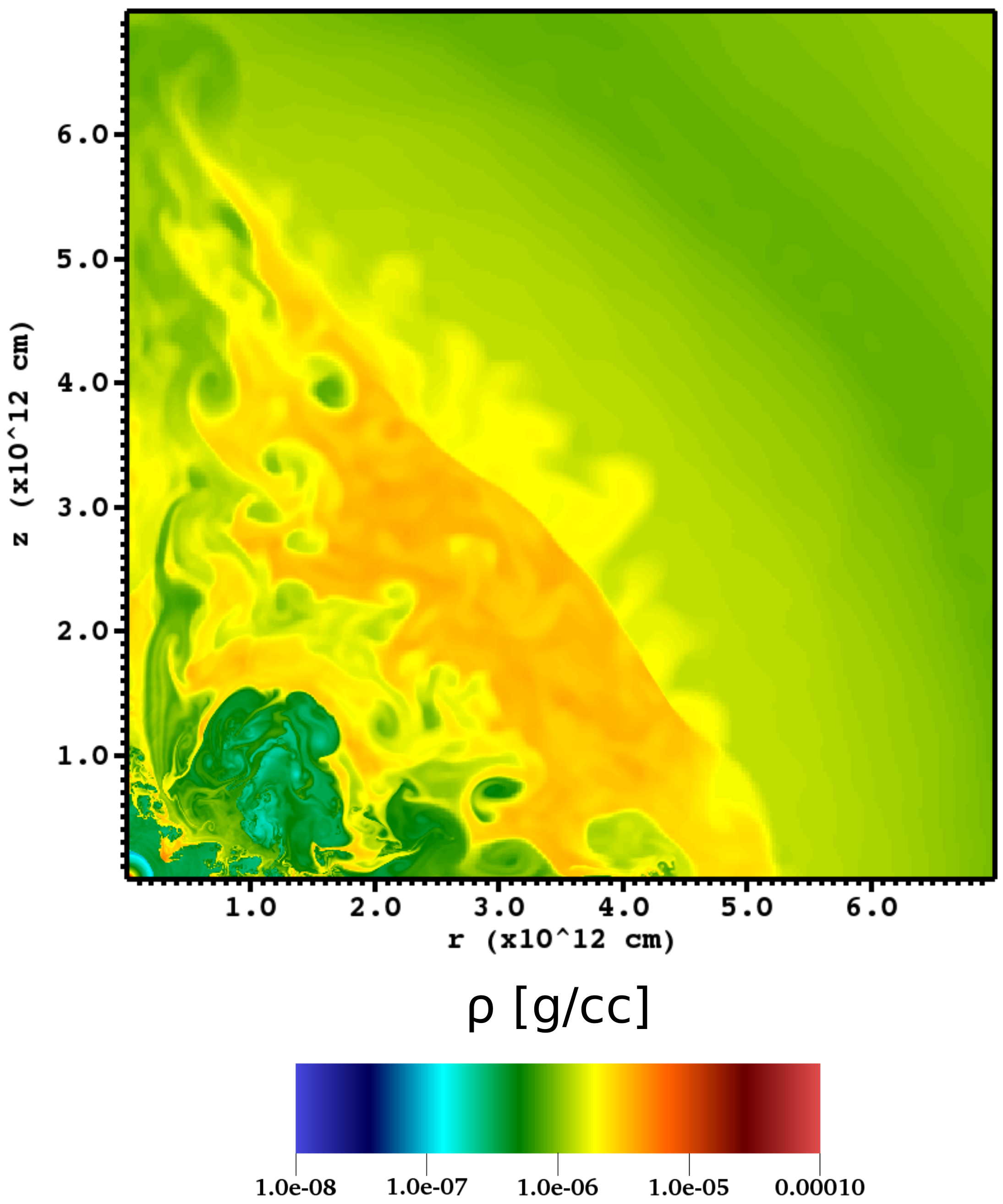} & 
\includegraphics[width=0.5\columnwidth]{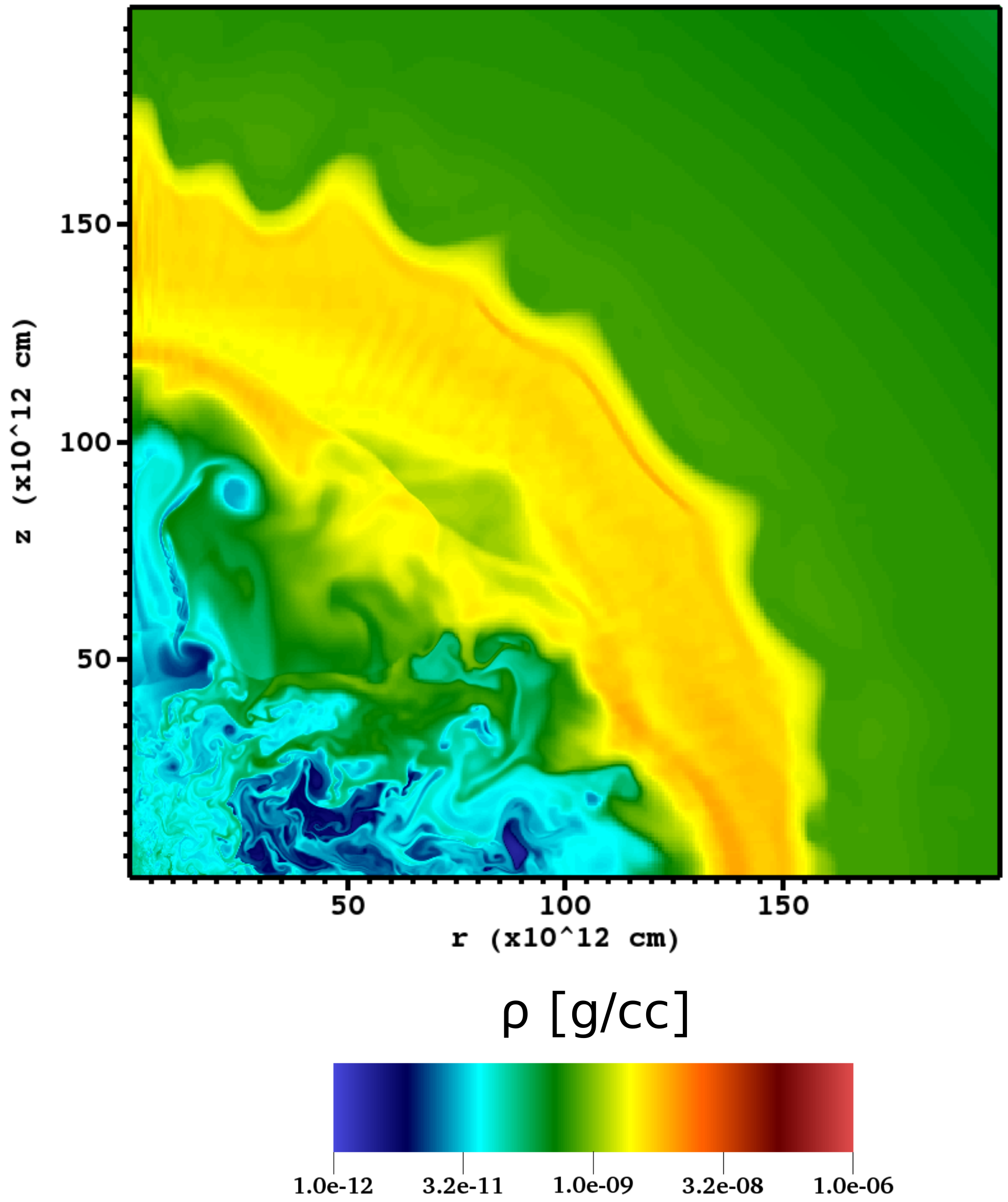} \\ 
\end{tabular}
\end{center}
\caption{Z12 (left) and Z60 (right) explosions at $t \sim$ 67,800 sec and $1.43 \times 10^6$ sec, respectively.  In both cases most of the ejecta is beyond $5r_*$ and is now in homologous expansion.}
\vspace{0.2in}
\label{fig:homol}
\end{figure}

\begin{figure}
\begin{center}
\begin{tabular}{cc}
\includegraphics[width=.50\columnwidth]{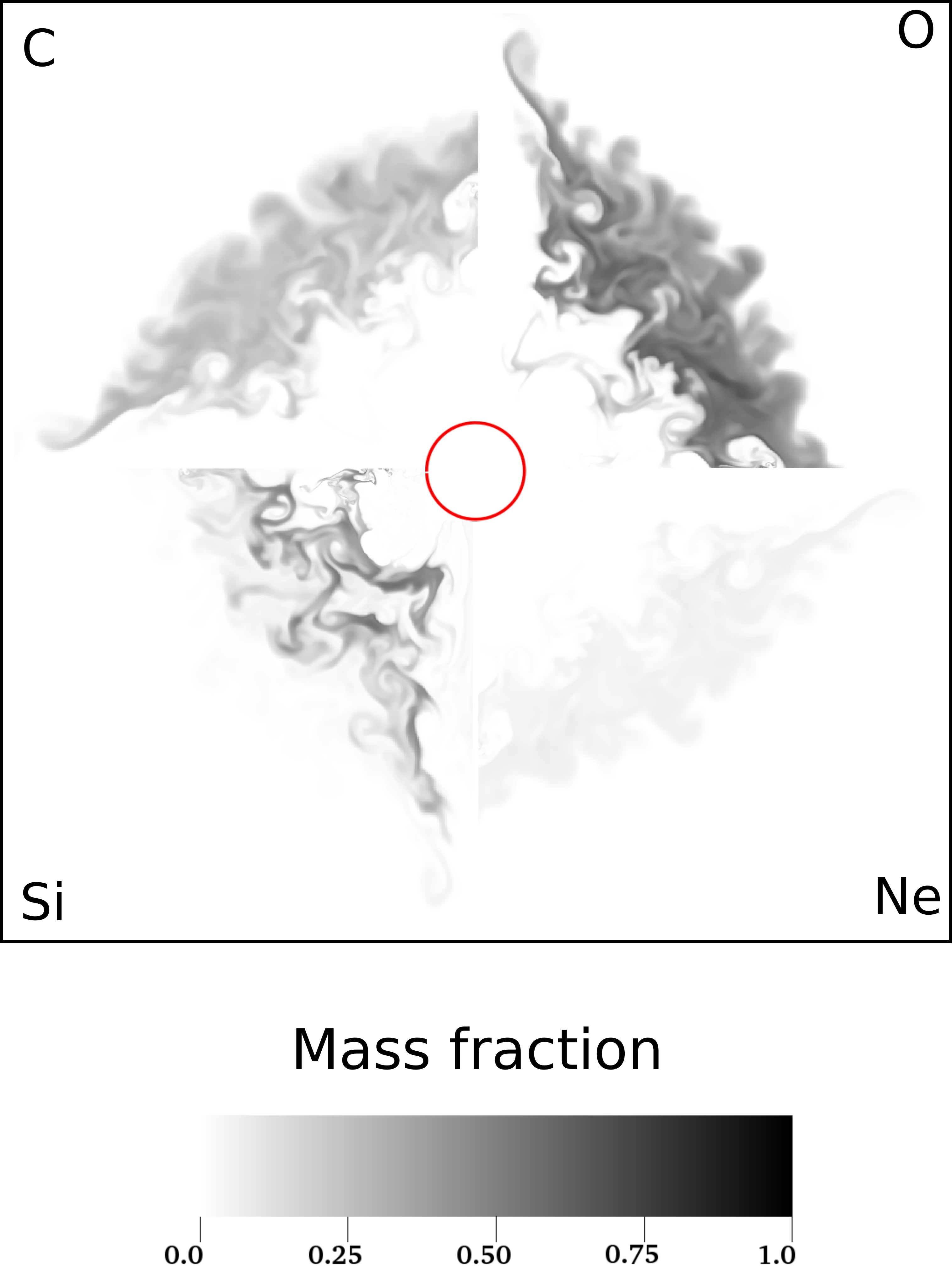} & 
\includegraphics[width=.50\columnwidth]{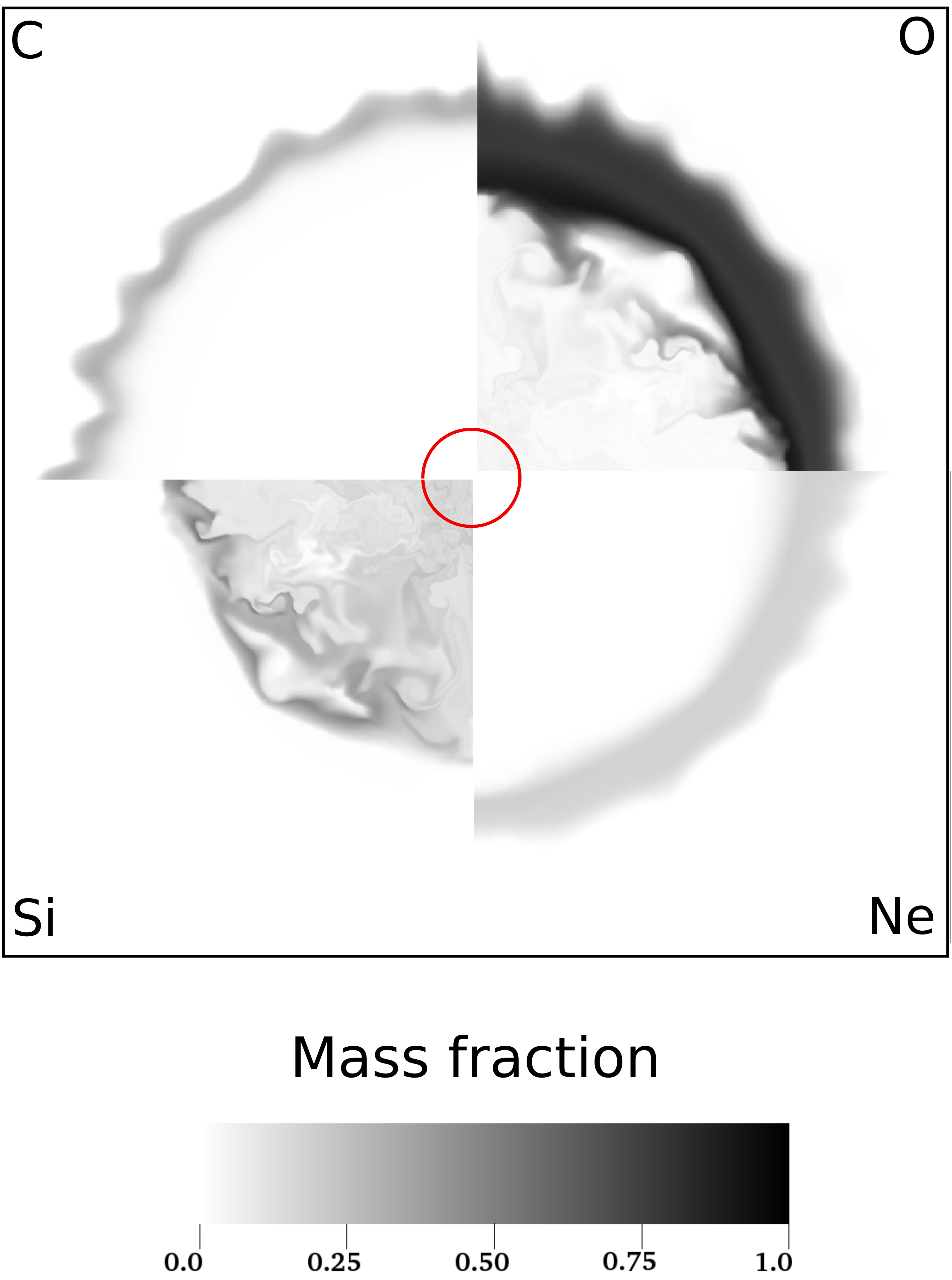} \\ 
\end{tabular}
\end{center}
\caption{Spatial distribution of the metals in the Z12 (left) and Z60 (right) SNe when their ejecta reach $\sim 5r_*$.  In each figure the red circle represents the radius of the progenitor star. In Z12, elements above \Si\ mostly fall back onto the compact remnant and the yields are dominated by \Cx\ and \Ox.  The yields in Z60 are similar to those of Z12 but \Cx\ and \Ox\ are not as mixed.}
\label{fig:metals}
\end{figure}

We show the distribution of the metal-rich ejecta at $r \gtrsim 5r_*$ in Figure \ref{fig:homol}.  Fallback is now essentially complete, and the ejecta expands homologously with an internal energy that is less than 10\% of its kinetic energy.  We show the distribution of various metals in the ejecta at this stage in Figure \ref{fig:metals}.  Strong mixing of \Cx\ and \Ox\ is visible in Z12 but the shells are more stratified in Z60.  Most of the mixing in Z60 occurs in the central regions and is due to fallback. In both cases a shell of carbon and oxygen is ejected but elements heavier than silicon mostly fall back to the compact remnant.  

\subsection{Why Mixing Occurs} 

\begin{figure}
\begin{center}
\includegraphics[width=\columnwidth]{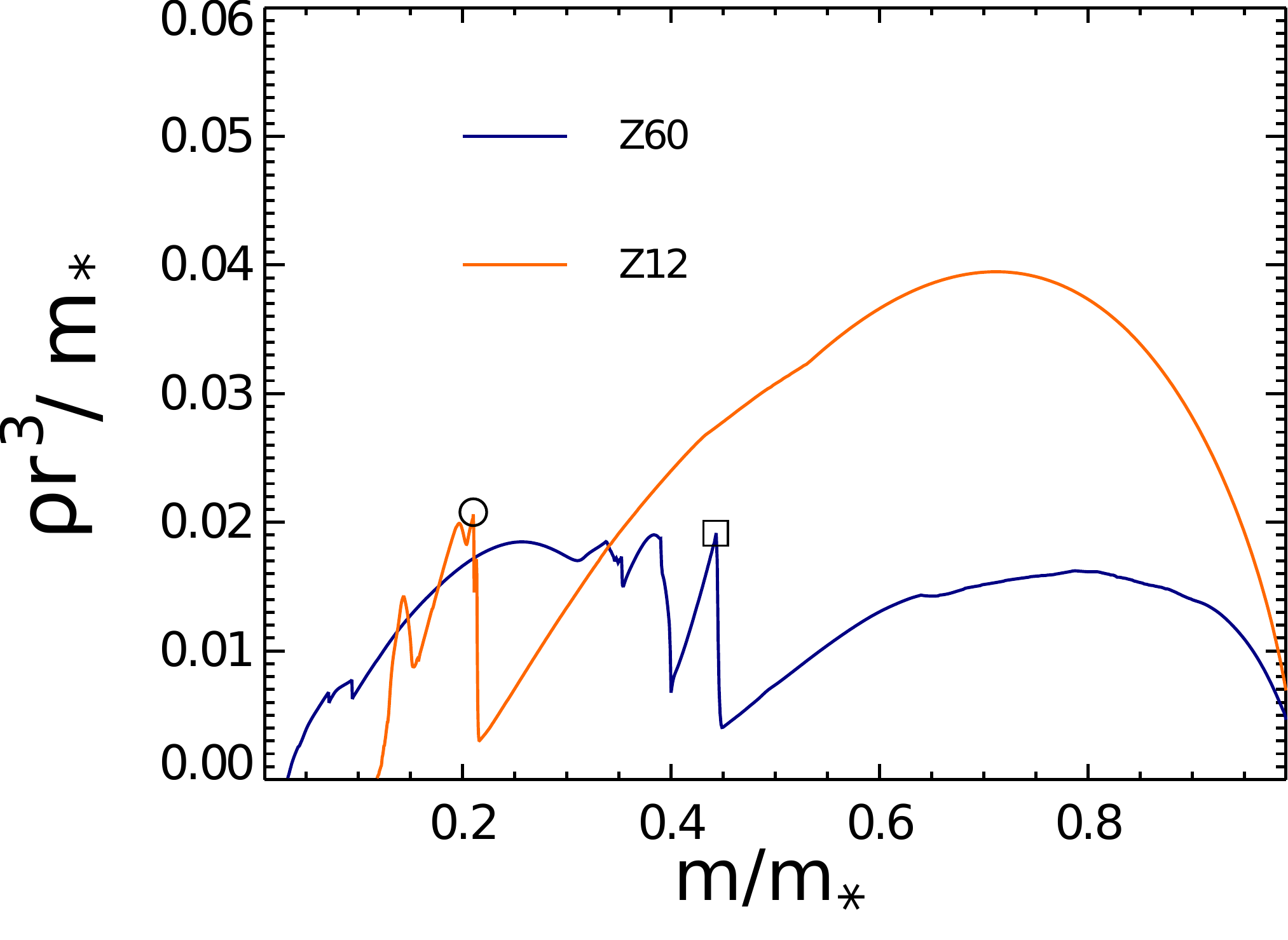} 
\caption{Internal structures of the Z12 and Z60 stars. The y-axis shows $\rho r^3$ normalized by the mass of the star, $m_*$, and the x-axis is the normalized mass coordinate. The circle and box mark the position of the shock. The broad peak at $m/m_* \sim 0.7$ corresponds to the hydrogen envelope of the star, which is more prominent and extended in Z12 than Z60. \label{fig:rhor3}}
\end{center}
\end{figure}

As noted above, mixing occurs at the earliest times and at the greatest depths during fallback, when the gravitational pull of the NS or BH arrests the expansion of the innermost ejecta, which are at relatively low velocities.  The relative orientations of the gravitational field and the acceleration of the ejecta satisfy the RT criteria and instabilities form.  Instabilities appear later as the shock propagates through the star.  As the shock plows up the star it decelerates, and a reverse shock forms and detaches from it if the structure of the star is $\rho \propto r^{w}$, where $w > -3$. The RT criteria are again satisfied and instabilities form. Consequently, an increasing $\rho r^3$ is a good indicator of where reverse shocks and RT instabilities can arise.  We plot $\rho r^3$ normalized by the mass of the star as a function of radius at the start of our \CASTRO{} runs in Figure \ref{fig:rhor3}.  In Z12, $\rho r^3$ is an increasing function over more of the mass coordinate of the star than in Z60 and one would therefore expect more mixing during the explosion, which is indeed the case. 

\subsection{Elemental Yields}

\begin{figure}
	\begin{center}
		\includegraphics[width=\columnwidth]{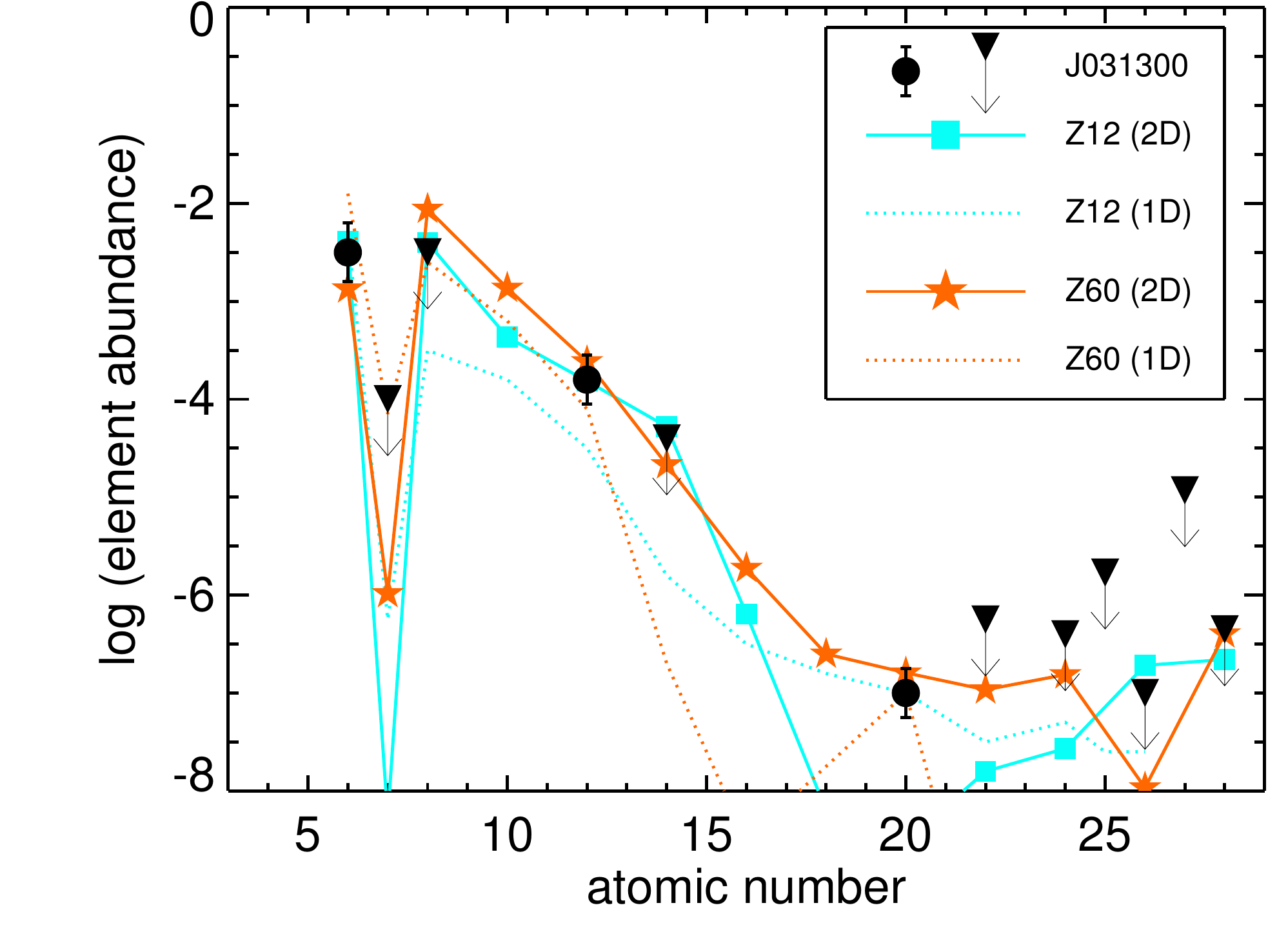} 
		\caption{2D \CASTRO\ and 1D \KEPLER\ yields for the Z12 and Z60 SNe with observed values (black circles) and limits (inverted black triangles) for J031300. Mixing during fallback, which is absent in 1D models, produces yields that are a better fit to those of J031300 at $Z > 20$.  Most of the individual elemental yields of the Z60 explosion are a better match to those of J031300 than those of the Z12 SN. \label{fig:yield}}
	\end{center}
\end{figure}

We summarize our yields in Table~\ref{tbl:iso1} and compare them with previous 1D models of the same explosions in Figure~\ref{fig:yield}. The 1D yields were obtained by choosing a radius below which ejecta is assumed to fall back onto the central object. Such 'mass cuts' are crude because any element below the cut in radius by construction cannot appear in the yields.  It is clear from Figure~\ref{fig:yield} that mixing in weak SNe can circumvent these cuts because the 2D models generally produce higher yields than the 1D explosions, particularly for isotopes with $Z \geq$ 20 deep in the ejecta.  Since even tiny amounts of dredged up material ($\leq 10^{-3} \Ms$) can change the abundances of the ejecta, it is clear that dynamical instabilities due to fallback are essential to computing realistic SN yields.  Z60 produces greater yields of heavy elements than Z12 relative to the mass of the progenitor star.

Abundances observed in J031300 for \Cx, \Mg, \Ca\ and upper limits on other isotopes are also plotted in Figure~\ref{fig:yield}. The 1D and 2D models of both SNe produce yields that are in reasonable agreement with \Cx, \Mg, \Ca, but the 2D yields are generally closer.  The differences between the 1D and 2D abundances are more pronounced at $Z \geq$ 20 because of mixing at early times due to fallback.  This mixing produces small amounts of heavy elements beyond \Ca\ that could found in future, more sensitive observations of J031300.

\subsection{SN Luminosities}

Because they have massive hydrogen-rich envelopes, Z12 and Z60 are expected look like  Type~IIP SNe, whose light curves are characterized by an intense breakout peak followed by a long plateau and finally a sudden drop in luminosity. \citet{kasen2009} estimate the luminosity and duration of the plateau to be
\begin{eqnarray}
L_{50} &=&1.26 \times 10^{42} {E_{51}}^{5/6} {M_{10}}^{-1/2} {R_{0,500}}^{2/3} X_{\mathrm{He}}~\mathrm{erg}~ {\mathrm{s}}^{-1}  \nonumber \\
t_{\mathrm{p},0} & = &122\, {E_{51}}^{-1/4} {M_{10}}^{1/2} {R_{0,500}}^{1/6} X_{\mathrm{He}}^{1/2}~\mathrm{days}, 
\end{eqnarray}
where $L_{50}$ is the bolometric luminosity on the plateau measured 50 days after explosion,  $R_{0,500} = R_0/500 R_{\odot}$ and $R_0$ is the radius of the star at the time of explosion, $t_{\mathrm{p},0}$ is the plateau duration when no \Ni\ is included, $E_{51}$ is the explosion energy in units of 10$^{51}$ erg, $M_{10}$ is the mass of the star in units of 10 \Ms\ and $X_{\mathrm{He}}$ is the mass fraction of helium.  Although these equations exclude the effect of \Ni\ on the light curve this is not a serious limitation because neither of our models produces any. From these estimates we obtain a plateau luminosity of $3.45 \times 10^{41} \mathrm{erg~s^{-1}}$ and duration of 83 days for Z12 and a luminosity of $1.54\times 10^{41}~\mathrm{erg~s^{-1}}$ and duration of 141 days for Z60.  These luminosities apply to solar-metallicity progenitors whose internal structures are probably different than those of zero-metallicity stars of equal mass.  But these rough estimates suggest that the plateau phase for Z12 and Z60 would not be visible at $z \gtrsim$ 5.  The shock breakout transient is much brighter but is primarily composed of X-rays and hard UV that would mostly be absorbed by the neutral IGM at $z \gtrsim$ 6 \citep{tominaga2011}.

\section{Conclusion}

Our numerical models confirm that the unusual abundance patterns of the most metal-poor star found to date, J031300, can be explained by mixing and fallback in weak Pop III SNe.  In particular, the apparent lack of iron in J031300 can be explained by the fallback of most of the iron group elements onto the NS or BH during the explosion.  Multidimensional mixing and fallback produce better fits to the chemical abundances of J031300 than 1D explosions of the same stars and energies.  The low explosion energies of these stars are similar to those of the unusually dim SNe that have recently been discovered in large numbers in the local universe \citep[e.g.,][]{Zam03,Pas07}.  Z60 produces chemical yields that are in slightly better agreement with those of J031300 than Z12, in which there is more violent mixing due to its internal structure.  There are apparent degeneracies in the elemental yields of Pop III SNe, given that explosions of progenitors with quite disparate masses can produce similar chemical abundances.  Other codes can also produce somewhat different yields for the same SNe due to differences in hydro scheme and reaction network \citep[see, e.~g.,][]{marass2014}.

The mixing in our 2D models should be taken to be a lower limit because they omit mixing due to asymmetries in the central engine.  These asymmetries, which \citet{Jak12} have shown can form before the launch of the shock \citep[and in some cases impart natal kicks to the compact remnant;][]{wf12}, might expel more iron group elements from the star than in our simulations.  Furthermore, while our 2D models produce reasonable fits to the chemical abundances in J031300 they do not explain the dilution factors with which those elements appear in the star, which are due to mixing of the SN remnant with the IGM on much larger scales.  When the ejecta plows up roughly its own mass in the halo a new reverse shock can form, driving more mixing \citep{wet08b}. Later, there could be a third stage of mixing when the ejecta collides with the dense shell of the relic H II region of the star at 100 - 200 pc \citep{wan04}, perhaps cooling and fragmenting it \citep{ritter2015,sluder2015}.  These fragments can then collapse into new stars via metal, molecule and dust cooling. Capturing all these processes with high resolution cosmological simulations is necessary to match both abundances and metallicities of any given metal-poor star to a specific SN.  It should be noted that second-generation, metal-poor stars could form in more complicated enrichment scenarios, such as ejecta from a CC SN colliding with a nearby halo, mixing with it, and triggering new star formation \citep{cr08,smith2015,chen16a}. Thus, multiple enrichment channels may have to be tested to tie the abundances of extremely metal-poor stars to specific Pop III SNe.

In our models we ignore the effects of jets and radiation from the NS and BH on fallback.  In principle, radiation from the compact remnant may decrease or reverse accretion.  If there is a strong jet it could drive out infalling material and burn it to higher atomic numbers, both of which could also alter the final yields of the explosion \citep[see, e.g.,][]{chen16b}.  Also, because there is no rotation in our models, accretion onto the central remnant is mostly spherical.  If rotation were present, fallback could collapse into a disk.  

The low-energy SNe in our study are likely more common than the highly energetic (and asymmetric) explosions invoked in the past to explain the unusual abundances of hyper metal-poor stars \citep[e.g.,][]{Iwamoto2005}.  These rare events require special conditions in order to occur, such as rapid rotation.  In contrast, the new SN factories such as the Palomar Transient Factory \citep[PTF;][]{ptf1} and the Panoramic Survey Telescope and Rapid Response System \citep[Pan-STARRS;][]{panstarrs} are discovering new, previously unknown classes of dim events at an unprecedented rate, with more soon to follow with the construction of the Large Synoptic Survey Telescope \citep[LSST;][]{lsst}.  Weak SNe in the local universe may therefore soon provide important clues about the chemical enrichment of the primordial universe. Unfortunately, the low luminosities of these events would preclude their direct detection at high redshift by the {\it James Webb Space Telescope} ({\it JWST}), the Wide-Field Infrared Space Telescope (WFIRST), {\em Euclid} and the coming generation of 30-40 meter class ground-based telescopes. The only traces of these ancient explosions will be in the fossil abundance record.

\begin{table*}
	\centering
	\caption{Masses of ejected elements and the compact remnant}
	\label{tbl:iso1}
	\begin{tabular}{lcccc} % four columns, alignment for each
		Model &  2D & 1D & 2D & 1D  \\  \hline
		Yields  &  Z12 & Z12 & Z60 & Z60 \\
		&   {\Ms} & {\Ms}  &   {\Ms} & {\Ms}     \\
		\hline
		\Hy &  6.51    & 6.41  & 18.84& 18.49   \\
		\He     &    3.98  & 3.94  &   19.17 &   16.86 \\
		\Cx      &   $4.91\times 10^{-2}$  &   $1.67\times 10^{-2}$   &   1.28 &   $1.71\times 10^{-2}$  \\ 
		\Nx      &   $7.40\times 10^{-8}$  &   $5.71\times 10^{-7}$  &  $1.56\times 10^{-4}$&  $2.68\times 10^{-5}$    \\ 
		\Ox     & 	  $6.38\times 10^{-2}$ &    $2.55\times 10^{-3}$   & 11.14& $7.88\times 10^{-3}$  \\
		\Ne     &  	$8.63\times 10^{-3}$ &  $3.12\times 10^{-4}$ &2.17 & $4.55\times 10^{-4}$ \\
		\Mg    &    $3.68\times 10^{-3}$  &  $3.28\times 10^{-5}$ &  $4.63\times 10^{-1}$ & $2.21\times 10^{-5}$  \\
		\Si    &   $1.47\times 10^{-3}$  &  $1.68\times 10^{-6}$   &  $4.77\times 10^{-2}$  & $6.06\times 10^{-8}$   \\
		\Sx &  $2.05\times 10^{-5}$  &  $1.53\times 10^{-7}$   & $4.78\times 10^{-3}$ & $8.98\times 10^{-10}$   \\
		\Ar     &    $2.28\times 10^{-7}$  &   $1.78\times 10^{-8}$   &  $7.15\times 10^{-4}$ & $2.99\times 10^{-11}$  \\
		\Ca     &     $1.21\times 10^{-7}$ &  $1.10\times 10^{-8}$   &  $5.11\times 10^{-4}$ & $3.54\times 10^{-9}$    \\
		\Ti     & 	$6.97\times 10^{-7}$  &  $1.04\times 10^{-10}$   & $3.77\times 10^{-4}$ & $5.61\times 10^{-13}$   \\  
		\Cr     &   $1.30\times 10^{-6}$ &   $1.23\times 10^{-9}$  &$5.87\times 10^{-4}$ & $1.27\times 10^{-13}$   \\
		\Fe     &    $9.95\times 10^{-6}$  &    $3.97\times 10^{-8}$  & $4.45\times 10^{-5}$ & $5.88\times 10^{-14}$    \\                            
		\Ni  &   $1.23\times 10^{-5}$  &   $1.12\times 10^{-9}$   &$1.79\times 10^{-3}$ & $4.54\times 10^{-14}$  \\	
		\hline
		ejecta   &  10.62  &  10.37  &53.13&  35.38 \\
		compact remnant &  1.78 (1.46+0.32)$^{\ast}$ &  2.03 &  6.87 (1.91+4.96)$^{\ast}$  & 24.62   \\
		total  & 12.40 & 12.40  &   60.00  & 60.00  
	\end{tabular}	
      \caption{Note: 1D models use artificial mass-cut and mixing to determine the yields. Therefore, the amount of ejecta  in 1D can be different from those in 2D. For the 1D Z60 model, almost the entire helium core is assumed to fallback to the central compact remnant. $^{\ast}$: (neutron star + fallback)   }

\end{table*}

\section*{Acknowledgements}

We thank the anonymous referee, whose comments and suggestions improved the quality of this paper. We also thank Ann Almgren and Weiqun Zhang for help with \CASTRO{} and Masaomi Tanaka, and Tom Jones for many useful discussions. K.C. acknowledges the support of EACOA Fellowship from the East Asian Core Observatories Association and the hospitality of Aspen Center for Physics, which is supported by National Science Foundation grant PHY-1066293.  Work at UCSC has been supported by an IAU-Gruber Fellowship, the DOE HEP Program (DE-SC0010676) and the NASA Theory Program  (NNX14AH34G). A.H. was supported by an ARC Future Fellowship (FT120100363). D.J.W. was supported by the European Research Council under the European Community's Seventh Framework Programme (FP7/2007-2013) via the ERC Advanced Grant "STARLIGHT: Formation of the First Stars" (project number 339177).  V.B. was supported by NSF grant AST-1413501.  \CASTRO{} was  developed through the DOE SciDAC program by grants DE-AC02-05CH11231, and DE-FC02-09ER41618. The numerical simulations were performed at the National Energy Research Scientific Computing Center (NERSC) and the Center for Computational Astrophysics (CfCA) at the National Astronomical Observatory of Japan (NAOJ).

\newpage

%\begin{figure}
%	\begin{center}
%		\includegraphics[width=0.8\columnwidth]{f3} 
%		\caption{Evolution of fluid instabilities in Z12 before and after shock breakout.  Each panel is $2 \times 
%			10^{11}$ cm on a side.  Panels (a) - (d) show the instabilities at 0, 340, 1200, and 2300 sec.  Fluid 
%			instabilities are driven by fallback as well as the reverse shock.  They form at the inner and outer 
%			boundaries of the carbon/oxygen burning shells.  \label{fig:z12evol}}
%	\end{center}
%\end{figure}

% \begin{figure}
%	\begin{center}
%		\includegraphics[width=0.5\columnwidth]{f5} 
%		\caption{Fluid instabilities in Z12 as the shock approaches the grid boundaries at $t\sim$ 44,000 sec.  
%			The reverse shock has reached the center of the star and mixing is now complete, although fallback 
%			continues. 
%			\label{fig:z12break}}
%	\end{center}
%\end{figure}

%\newpage

% Example table

%\bibliographystyle{mn2e}
%\bibliography{Chen_K}

\end{document}